\documentclass[aps,english,prb,reprint,superscriptaddress,onecolumn]{revtex4-1}

\usepackage{graphicx}
\usepackage{amssymb}
\usepackage{amsmath}
\usepackage{babel}
\usepackage{color}
\usepackage{multirow}
\usepackage[T1]{fontenc}
\usepackage{siunitx}
\usepackage{soul}
\usepackage{url}

\begin{document}
\title{\textbf{Emergent Antiferromagnetic Behavior in EuCl$_3$ Monolayer: A Comprehensive First-Principles Study Including Hubbard-SOC Interplay}}

\author{Nurcan Kutluk Kalkan}
\affiliation{Physics Department, Adnan Menderes University, 09100 Aydin,
Turkey}

\author{Ilkay Ozdemir}
\affiliation{Physics Department, Adnan Menderes University, 09100 Aydin,
Turkey}
\affiliation{Department of Physics, University of Antwerp, Groenenborgerlaan 171, B-2020 Antwerp, Belgium}

\author{Yusuf Y\"{u}ksel}
\affiliation{Physics Department, Faculty of Science, Tinaztepe Campus, Dokuz Eylul University, 35390 Izmir, Turkey.}

\author{\"{U}mit Ak{\i}nc{\i}}
\affiliation{Physics Department, Faculty of Science, Tinaztepe Campus, Dokuz Eylul University, 35390 Izmir, Turkey.}

\author{Johannes V. Barth}
\affiliation{Physics Department E20, Technical University of Munich (TUM), James Franck Strasse1, 85748, Garching (Germany)}

\author{Ethem Akt\"urk}
\email{ethem.akturk@adu.edu.tr}
\affiliation{Physics Department, Adnan Menderes University, 09100 Aydin,
Turkey}
\affiliation{Physics Department E20, Technical University of Munich (TUM), James Franck Strasse1, 85748, Garching (Germany)}

\date{\today}

\begin{abstract}

We present a comprehensive study on the electronic and magnetic properties of the EuCl$_3$ monolayer using first-principles calculations.  By taking into account the spin-orbit coupling (SOC) and the Hubbard effects,  we elucidate the influence of these interactions on the structural properties of EuCl$_3$ monolayer.  Comparing the lattice parameters obtained with the PBE+SOC and PBE+SOC+Hubbard effects reveals an increase in the lattice parameters when including the Hubbard effect.  In the absence of the Hubbard interaction, the magnetic ground state of the EuCl$_3$ monolayer exhibits a preference for a ferromagnetic (FM) configuration as determined by the PBE+SOC calculations.  However, the introduction of the Hubbard parameter leads to a shift in the magnetic ground state preference towards an antiferromagnetic (AFM) N\'{e}el state.  Based on the calculated energy values, Monte Carlo simulations are carried out to determine the N\'{e}el temperature $(T_N)$.  Our simulations yield a N\'{e}el temperature of \SI{390}{\kelvin} for the EuCl$_3$ monolayer, indicating the temperature at which the transition from a paramagnetic phase to an antiferromagnetic phase occurs.  These results highlight the importance of incorporating the SOC and the Hubbard effect in accurately describing the electronic and magnetic properties of the EuCl$_3$ monolayer. Our results contribute to a deeper understanding of the fundamental physics underlying the behavior of this intriguing two-dimensional material and provide insights into its potential applications in spintronics and magnetic devices.

\end{abstract}

\maketitle


\section{Introduction}

Spintronics research has experienced significant growth recently, particularly in the exploration of two-dimensional (2D) van der Waals (vdW) structures \cite{zhang2019, klein2018, khan2020, wang2004, feng2020}. These structures have emerged as a prominent area of investigation, enabling fundamental studies of spin-based devices and facilitating various spin injection and spin-polarized transport theories \cite{webster2018}. 2D vdW magnets exhibit intrinsic ferromagnetic (FM) and antiferromagnetic (AFM) ground states, even at atomic layer thicknesses, thus presenting exciting opportunities for materials science and the development of spin-related applications. The layered structure of these 2D vdW magnetic systems offers advantages such as atomic layer divisibility and magnetic anisotropy, effectively suppressing spin fluctuations \cite{li2019}. Consequently, they provide an ideal platform for both theoretical and experimental exploration of phase transitions at the 2D boundary. Despite these advancements, two major challenges persist in the realm of 2D structures. Firstly, the identification of materials that demonstrate stability under ambient conditions, possess long-term operational suitability, and exhibit sufficient mechanical strength remains essential \cite{cortie2020}. Secondly, it is crucial to elevate the magnetic phase transition temperatures above room temperature. Among the significant properties of magnetic materials, the critical temperature $(T_c)$ holds paramount importance as it represents the temperature at which magnetic order is lost. Achieving a high $T_c$, surpassing room temperature, becomes a critical requirement for effective implementation of FM spintronic devices \cite{li2019}.

The Mermin-Wagner theorem first established constraints on the presence of long-range magnetic order in 2D systems at finite temperature \cite{mermin1966absence}. According to this theorem, the spontaneous breaking of continuous symmetries, such as the rotational symmetry or the U(1) spin symmetry,  is prohibited in 2D systems at any finite temperature.  This would suggest an absence of long-range magnetic order in 2D materials. However, the conventional understanding was challenged in 2017 with the discovery of ferromagnetism in monolayer CrI$_3$ \cite{huang2017layer}. Key to understanding this apparent contradiction is the interplay of several factors, including the inherent magnetic anisotropy due to the strong spin-orbit coupling and the role of defects or lattice distortions.  Following this experimental discovery, demonstrating the ability of a 2D material to maintain magnetization at a non-zero temperature,  2D magnetic materials have been at the forefront of both theoretical \cite{lado2017origin, xu2018interplay, gibertini2019} and experimental \cite{gong2017discovery, deng2018gate, bonilla2018strong} scientific investigations.  In the case of the CrI$_3$ monolayer,  the ferromagnetism appears to arise from a combination of robust magnetic anisotropy and the suppression of thermal fluctuations due to quantum confinement \cite{huang2017layer}.   The existence of an out-of-plane easy-axis magnetic anisotropy in CrI$_3$, originating from interlayer magnetic coupling, helps stabilize the ferromagnetic order in the monolayer despite the limitations imposed by the Mermin-Wagner theorem.  Further studies, both experimental and theoretical, have also unveiled the presence of magnetism when other chalcogen atoms are incorporated.  For instance, bulk CrBr$_3$ is an easy-axis ferromagnet with a Curie temperature ($T_c$) of \SI{33}{\kelvin} \cite{gong2019two}, and it retains ferromagnetism even at monolayer thickness, exhibiting pronounced out-of-plane anisotropy \cite{kim2019, bacaksiz2021}. Conversely, bulk CrCl$_3$ is an easy-plane A-type antiferromagnet \cite{gong2019two}, while investigations on few-layer samples have highlighted an enhancement of the antiferromagnetic interlayer exchange compared to bulk crystals \cite{klein2019, serri2020}.

On the other hand, rare-earth metal elements, especially gadolinium (Gd) and europium (Eu), possess half-filled and highly localized 4f subshells, leading to large magnetic moments and strong spin-orbit coupling (SOC) compared to transition metal elements \cite{tokmachev2019lanthanide, jiang2021recent}. As a result, due to their unique electronic and magnetic properties, rare-earth-based 2D materials, especially monolayers composed of Gd or Eu, have garnered increasing attention. These materials provide an exciting platform for investigating novel phenomena and advancing technologies such as spintronics, quantum computing, and magnetic storage devices. Gd-based monolayers, renowned for their high Curie temperatures near room temperature, exhibit robust magnetic properties, positioning them as promising candidates for spintronic applications \cite{Law2010, Shen2009, Pecharsky1997}. Moreover, it has been reported that Eu doping in two-dimensional materials can enhance their electronic properties, magnetism, and critical temperature \cite{Assadi2021, Chakraborty2021, Zhao2015, Franco2014, Masago2020}. Building upon these findings, our study focuses on investigating the electronic and magnetic properties of the EuCl$_3$ monolayer. Our calculations incorporated the Hubbard effect along with SOC to better define the structure, considering previous studies that have highlighted the significance of this addition \cite{Kishore2016, Durajski2020, Kirchner2021, Das2021, Ersan2019, Abdullahi2021}. Comparison of lattice parameters obtained under the PBE+SOC and PBE+SOC+Hubbard effects revealed that the inclusion of the Hubbard effect led to an increase in lattice parameters. While the magnetic ground state of EuCl$_3$ favored a ferromagnetic (FM) configuration in the PBE+SOC calculations, the introduction of Hubbard parameter shifted the preference to an antiferromagnetic (AFM) N\'{e}el state. Magnetic anisotropy energies were obtained by exploring different spin orientations within the structure. Based on the resulting energy values, Monte Carlo simulations were performed, determining a N\'{e}el temperature $(T_N)$ of \SI{390}{\kelvin} for EuCl$_3$ monolayer.

\section{Computational Methods}

First-principles calculations were carried out using the spin-polarized density functional theory (DFT) implemented in the Vienna Ab initio Simulation Package (VASP) \cite{vasp1,vasp2}. We described the ion-electron interactions by using projector augmented wave (PAW) potentials \cite{paw1,paw2}. The Perdew-Burke-Ernzerhof (PBE) functional \cite{pbe}, which is based on the generalized gradient approximation (GGA), was employed to account for the exchange-correlation interactions. The electronic and magnetic properties of the system were determined by solving the Kohn-Sham equations self-consistently within this framework. The calculations were performed with a finite basis set using plane-wave expansions, and a cutoff energy of \SI{340}{\eV} was chosen to ensure convergence. To account for the van der Waals interactions, the DFT-D2 correction \cite{vdW-grimme} scheme was included. The Brillouin zone integration was performed using a $12\times12\times1$ mesh points in \textbf{k}-space within Monkhorst-Pack grid \cite{monkhorstpack}. To optimize the atomic positions and lattice constants, we employed the conjugate-gradient algorithm \cite{conjugategradient1,conjugategradient2}. The optimization process involved minimizing the total energy and forces to ensure convergence. A convergence criterion of \SI{e-05}{\eV} was set, ensuring that the energy difference between two consecutive steps reached a sufficiently low value. During the ionic relaxation, the maximum Hellmann-Feynman forces acting on each atom were constrained to be below \SI{0.01}{\eV/\angstrom}, guaranteeing stability within the system. Moreover, the maximum pressure exerted on the unit cell was maintained below \SI{1}{\kilo bar}. To handle the electronic occupations, we implemented the Gaussian-type Fermi-level smearing method with a smearing width of \SI{0.01}{\eV}. Visualization of the structures was achieved using the VESTA program \cite{vesta}.

All calculations, unless otherwise stated, incorporated the effects of spin-orbit coupling (SOC) and Hubbard interactions. The inclusion of SOC accounts for the strong interaction between the electron spin and its orbital motion, which is crucial for accurately describing the electronic and magnetic properties of the system. Additionally, the Hubbard effect, accounting for the on-site electron-electron interactions, was considered to better capture the electronic correlations within the system. To determine the appropriate effective Coulomb interaction parameter ($U_{eff}$) for the EuCl$_3$ monolayer, the linear response approach introduced by Cococcioni and De Gironcoli \cite{cococcioni2005linear} was employed by using the VASP program. Specifically, the effective U value for the EuCl$_3$ monolayer was found to be $U_{eff}$=\SI{8.9}{\eV}. (See the Supporting Information \cite{suppl} for more details.) Our obtained value is comparable with the findings from Pradip et al. \cite{pradip2016lattice}, who investigated the EuO crystal structure. In their study, they determined the $U$ value for the $f$-orbitals of Eu as \SI{8.3}{\eV} and the $J$ value as \SI{0.77}{\eV}. By incorporating SOC and the Hubbard effect, our calculations provide a comprehensive understanding of the electronic and magnetic properties of the EuCl$_3$ monolayer.

We calculated the work function $W$ for the EuCl$_3$ monolayer by the difference \cite{kittel}
\begin{equation}\label{equ1}
W=-e\phi-E_F
\end{equation}
where $-e$ represents the charge of an electron, $\phi$ denotes the electrostatic potential in the vacuum nearby the surface, and $E_F$ is the Fermi level, which corresponds to the electrochemical potential of electrons within the material. The term $-e\phi$ represents the energy of a stationary electron in the vacuum nearby the surface.

To investigate the magnetic properties of the EuCl$_3$ monolayer, we employed an atomistic spin Hamiltonian based on a classical Heisenberg model, which can be expressed by the following equation:
\begin{eqnarray}\label{equ2}
\nonumber
\mathcal{H}&=&-J_{1}\sum_{\langle
ij\rangle_{1}}\vec{\mu}_{i}.\vec{\mu}_{j} -J_{2}\sum_{\langle
ik\rangle_{2}}\vec{\mu}_{i}.\vec{\mu}_{k} -J_{3}\sum_{\langle
il\rangle_{3}}\vec{\mu}_{i}.\vec{\mu}_{l}\\
&-&\sum_{i}(k_{x}\mu^2_{ix}+k_{y}\mu^2_{iy}+k_{z}\mu^2_{iz}),
\end{eqnarray}
Here, $\vec{\mu}_{i}$ represents the classical magnetic dipole moment located at site $i$ within a honeycomb lattice of linear dimension $L=100$. $J_{1}$, $J_{2}$, and $J_{3}$ denote the exchange coupling between nearest, second-nearest, and third-nearest neighbor spins, respectively. We considered four distinct magnetic configurations, comprising one ferromagnetic state and three antiferromagnetic states. These magnetic configurations are illustrated in Fig. \ref{fig2}. To determine the values of the exchange coupling parameters, we calculated the total energies associated with each magnetic configuration and employed these calculated energies in the equations provided in the Supporting Information \cite{suppl}. The fourth term in Equation \ref{equ2} accounts for the contribution of single-ion anisotropy to the total energy per spin.

In order to delve deeper into the magnetic properties of the EuCl$_3$ monolayer, we carried out Monte Carlo simulations in addition to the DFT calculations. By combining these two computational techniques, we aimed to provide a comprehensive understanding of the magnetic behavior in EuCl$_3$ monolayer. The Monte Carlo simulations were based on the microscopic Hamiltonian, treating each spin as a classical vector with a magnitude of $\mid\vec{\mu}\mid$=\SI{6.0}{\micro_B}. The simulations utilized the Metropolis update scheme \cite{newman,binder} and incorporated periodic boundary conditions in each direction. To minimize statistical errors, we considered 50 different sample realizations at each temperature. Starting from a randomly oriented initial configuration at a temperature well above the critical temperature of the system, we gradually decreased the temperature in small steps while monitoring various thermal and magnetic properties, including the magnetic order parameter $\mathbf{O}=\frac{1}{L^{2}}\sum_{i=1}^{L^{2}}(-1)^{i}\vec{\mu}_{i}$ and the heat capacity $\mathbf{C}=\partial{\langle\mathcal{H}\rangle}/\partial{T}$.

\section{Results}

\begin{figure}[h]
\centering
\includegraphics[scale=0.3]{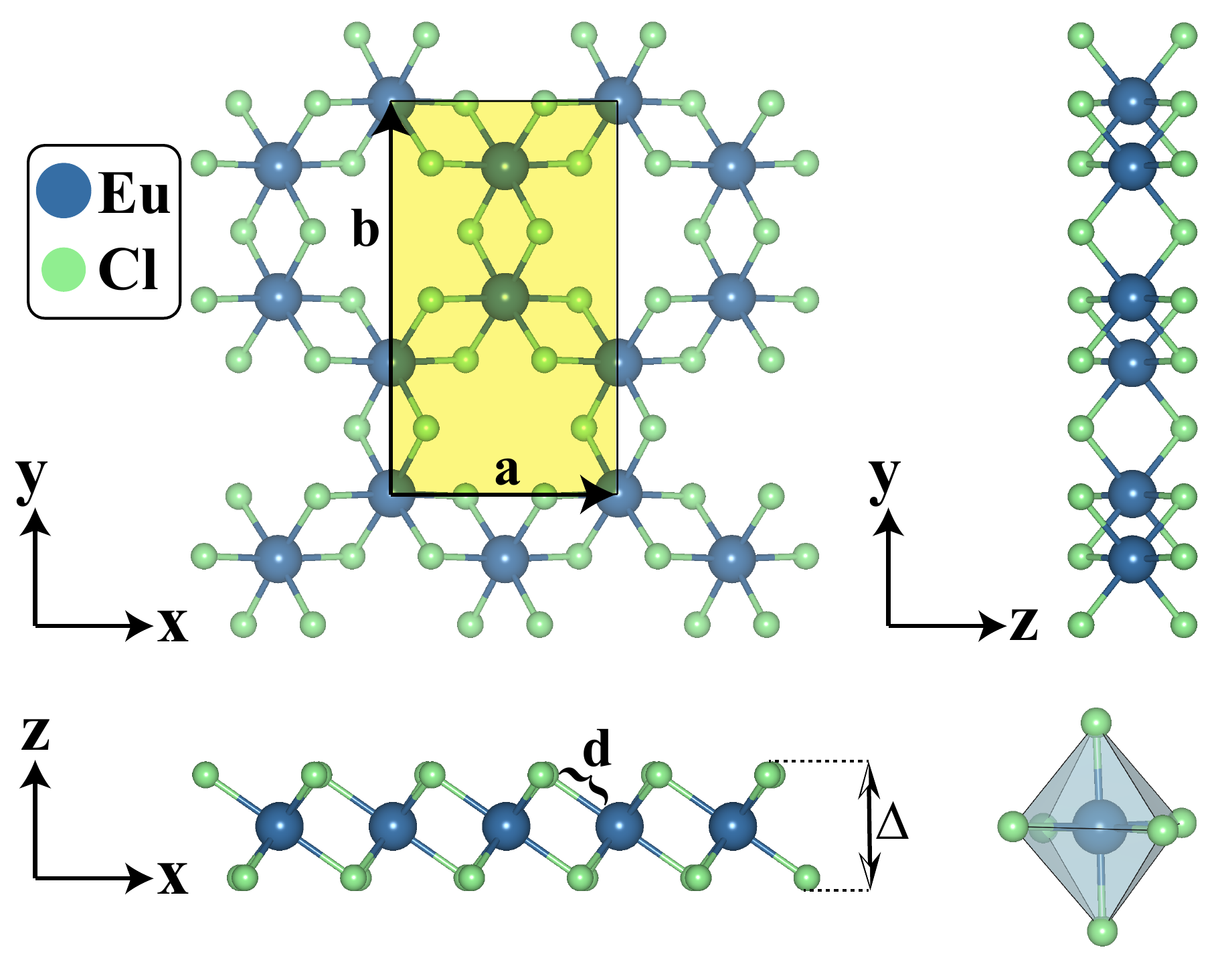} 
\caption{Crystal structure of the EuCl$_3$ monolayer, presented from three different views. The unit cell is highlighted with a yellow shaded area. The octahedral geometry is also given inset, which is depicted in blue shading.}
\label{fig1}
\end{figure}

EuCl$_3$ monolayer, examined in this study, exhibits a trigonal symmetry within the $P\bar{3}1m$ space group, similar to the well-known transition metal trihalides \cite{Ersan2019}. Its optimized crystal structure, as shown in Fig. \ref{fig1}, comprises a single layer of europium (Eu) atoms situated between two layers of chlorine (Cl) atoms. Within this crystal structure, each Eu atom is coordinated by six Cl atoms, resulting in an octahedral coordination geometry. The Eu atoms are arranged in a honeycomb lattice, with three neighboring Cl atoms bonded to each Eu atom from opposing sides. The Cl atoms form hexagonal layers positioned above and below the Eu layer, thus creating a structure with three-atomic-layer.

In the first step, we performed the optimization of the EuCl$_3$ monolayer using the standard PBE method, which provided a baseline structure. Subsequently, to take into account for the spin-orbit coupling (SOC) effect, we incorporated it into the calculation and further optimized the structure using the PBE+SOC method.  Finally, we enhanced the accuracy of our study by introducing the Hubbard effect, which considers the localized electrons in the $f$-orbitals of the Eu atom. This comprehensive approach, known as PBE+SOC+U calculation, enables a more thorough treatment of electronic interactions. In the latter case, we observed an expansion of the lattice parameters, in particular a \SI{2}{\percent} increase in the x-crystallographic direction and a \SI{0.1}{\percent} increase in the y-crystallographic direction. The calculated structural parameters are presented together in Table \ref{table1}.

\begin{table}[h]
\centering
\caption{Structural parameters of the EuCl$_3$ monolayer calculated using different methods (PBE, PBE+SOC, PBE+SOC+U): Lattice constants along the $x$-crystallographic direction ($\mid\vec{a}\mid$=a) and the $y$-crystallographic direction ($\mid\vec{b}\mid$=b); thickness parameter ($\Delta$); average bond lengths between neighboring Eu and Cl atoms ($d$); and average magnetic moment on the unit cell ($\mu$).}
\begin{tabular}{cccccc}
\hline \hline
method & a (\si{\angstrom}) & b (\si{\angstrom}) & $\Delta$ (\si{\angstrom}) & d (\si{\angstrom}) & $\mu$ ($\mu_B$)   \tabularnewline \hline
PBE        &6.64  &11.77  &3.11  &2.70  &24.00    \tabularnewline
PBE+SOC    &6.64  &11.77  &3.11  &2.70  &23.86    \tabularnewline
PBE+SOC+U  &6.77  &11.79  &3.06  &2.69  &23.79    \tabularnewline
\hline \hline
\label{table1}
\end{tabular}
\end{table}

\begin{figure}[h]
\centering
\includegraphics[scale=0.3]{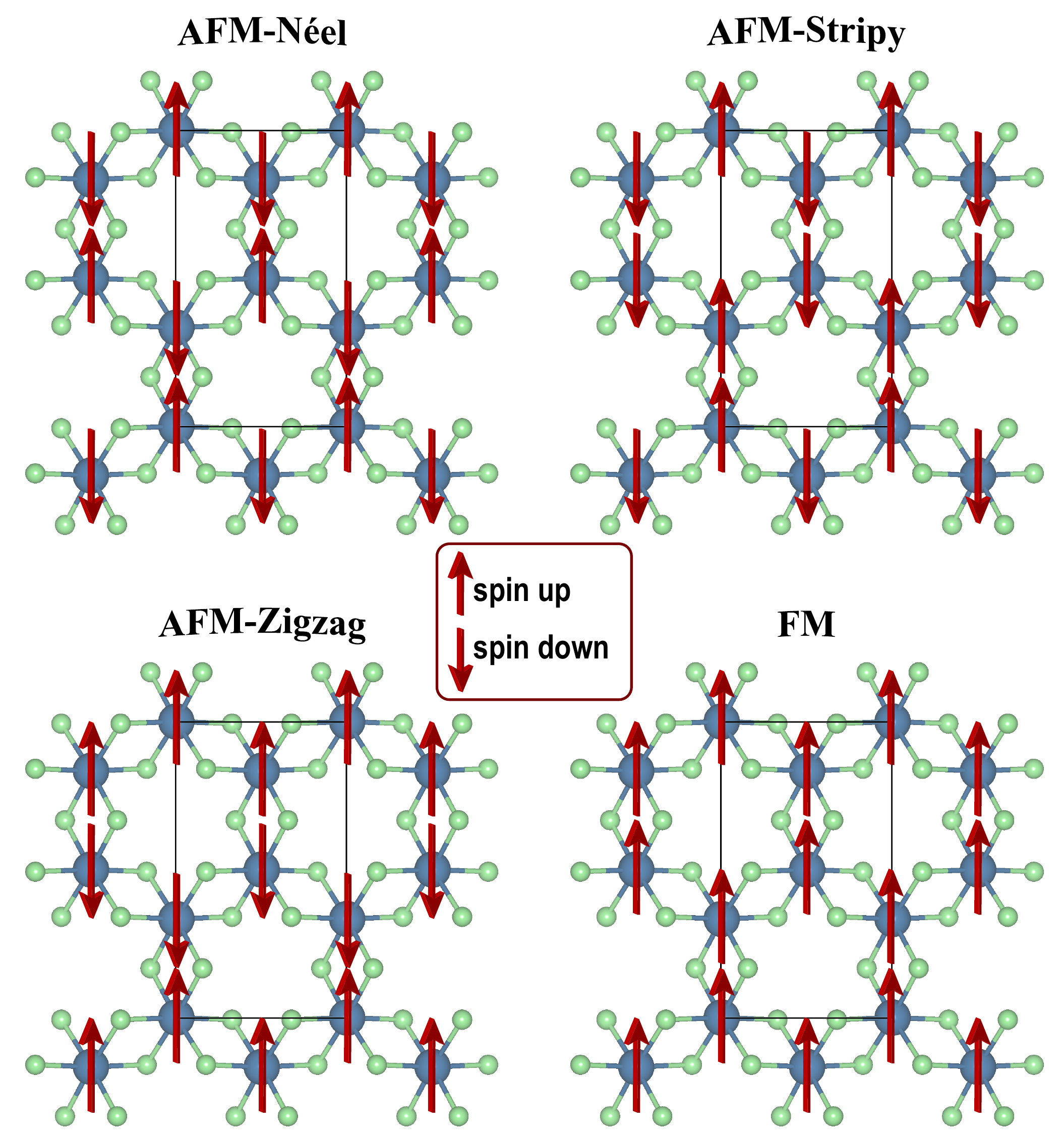} 
\caption{Magnetic configurations investigated for the EuCl$_3$ monolayer. Arrows pointing upwards indicate spin-up states, while arrows pointing downwards represent spin-down states.}
\label{fig2}
\end{figure}

After performing the structural optimization, we proceeded to analyze the ground state of the EuCl$_3$ monolayer in terms of magnetism.  In order to determine the most favorable magnetic ground state configuration, we examined four different spin configurations for the europium atoms, as depicted in Fig. \ref{fig2}. These configurations are labeled as ferromagnetic (FM), antiferromagnetic-N\'{e}el (AFM-N\'{e}el), AFM-Stripy, and AFM-Zigzag. Our results from the PBE+SOC calculations indicated that the EuCl$_3$ monolayer prefers a ferromagnetic configuration as its magnetic ground state.  However, upon incorporating the Hubbard effect through PBE+SOC+U calculations, the magnetic ground state preference shifted to an antiferromagnetic N\'{e}el state. We provided the relative energies of magnetic configurations with respect to the corresponding minimum energy configurations, both calculated without and with the inclusion of the Hubbard parameter, in the Supporting Information \cite{suppl}. The inclusion of the Hubbard term provides a more accurate treatment of the electron-electron interactions, particularly those involving the localized $f$-orbitals of the Eu atoms. The Coulomb repulsion within these orbitals promotes an antiferromagnetic alignment of the magnetic moments, resulting in cancellation of the net magnetic moment within the cell.  In the resulting AFM-N\'{e}el state, we calculated the magnitude of the magnetic moments on the magnetic Eu atoms as \SI{6.07}{\micro_B}. However, to simplify further magnetic calculations in Monte Carlo simulations, we rounded this value to \SI{6.00}{\micro_B}. This rounded value still captures the essential magnetic behavior of the system and ensures computational efficiency in subsequent simulations. These findings demonstrate the significant influence of both spin-orbit coupling and electron-electron interactions on the magnetic ground state of the EuCl$_3$ monolayer. The transition from ferromagnetic to antiferromagnetic behavior highlights the delicate balance between competing magnetic interactions in this system, which can be harnessed for potential applications in spintronics and magnetic devices.

In the AFM-N\'{e}el ground state of the EuCl$_3$ monolayer, we first examined the stability of its atomic structure. The monolayer exhibits a cohesive energy of \SI{3.50}{\eV} per atom, which indicates a strong binding between the atoms. To further assess its stability, we conducted dynamical stability analyses through phonon calculation, without taking into account the effects of SOC due to its computational cost. In the Supporting Information \cite{suppl}, we presented our findings regarding the phonon frequencies throughout the Brillouin zone, as well as the phonon projected density of states. Remarkably, we observed predominantly positive phonon frequencies throughout the entire Brillouin zone, with the exception of the region around the $Y$-point. However, it is important to note that despite the negativity observed around the $Y$-point, our analysis reveals that there is no density of states below the zero frequency level. This indicates that the EuCl$_3$ monolayer exhibits overall stability and is not prone to instability.

After achieving its stable ground state in the AFM-N\'{e}el configuration, we conducted a comprehensive investigation of the electronic properties of the EuCl$_3$ monolayer. To explore the electronic charge transfer mechanism within the system, we employed Bader charge analysis \cite{bader}. Our calculations demonstrated that each Eu atom donated \SI{2.01}{e}, resulting in a cumulative transfer of \SI{8.04}{} electrons that were equally shared among the Cl atoms. As a consequence, each Cl atom exhibited an excess charge of \SI{0.67}{e}. Further information regarding this charge transfer mechanism, the calculated charge density difference isosurfaces, can be found in the Supporting Information \cite{suppl}. These isosurfaces were obtained by subtracting the charge densities of isolated atoms from the entire system's charge density. The observed direction of charge transfer aligns with the concept of Pauling electronegativity, indicating that the electronegative Cl atoms tend to gain electrons while the Eu atoms, being less electronegative, donate electrons. For Cl, the Pauling electronegativity is known to be \SI{3.12}, whereas there are no reliable sources for Eu other than the one reported by Pauling in 1960 \cite{pauling1960nature} with the value falling within the range of \SIrange{1.1}{1.2}{}. Besides, we determined the work function of the AFM-N\'{e}el EuCl$_3$ monolayer by using the Equation \ref{equ1}. The work function, a fundamental concept in solid-state physics, represents the minimum thermodynamic work (i.e., energy) required to extract an electron from a solid to a point in the vacuum located immediately outside the solid surface. Our calculations yielded a work function of $W$=\SI{0.57}{\eV} for the AFM-N\'{e}el EuCl$_3$ monolayer. Furthermore, we performed electronic energy band structure calculations for the EuCl$_3$ monolayer, in the AFM-N\'{e}el ground state, using both the standard PBE functional and the HSE hybrid functional, which is known for its improved accuracy in predicting band gap values. Additionally, we generated the corresponding total and partial electronic density of states (PDOS) graphs through PBE functional.  All these plots have been presented in the Supporting Information \cite{suppl}. Our investigation of the electronic energy band diagrams under the PBE functional, calculated along the major symmetry directions of the Brillouin zone, demonstrates a semiconductor behavior with a band gap value of \SI{1.75}{\eV}. However, the HSE functional predicts the EuCl$_3$ monolayer to be an insulator, featuring a significantly wider band gap of \SI{4.99}{\eV}. Analyzing the partial DOS graph in detail, we observed that the main contribution to the valence bands around the Fermi level originates from Cl atoms, with a relatively minor contribution from Eu atoms. Furthermore, we found that the DOS of Eu atoms in the valence bands is primarily due to $d$-orbital electrons, while $f$-orbital DOSs remain nearly negligible. In contrast, the conduction bands feature a rather flat band around \SI{1.60}{\eV}, predominantly arising from the $f$-orbitals of Eu atoms. The presence of $f$-orbital electrons in the conduction band indicates a significant role of Eu's $4f$ electronic states in the conducting behavior of the EuCl$_3$ monolayer.

Next, we focused on investigating the magnetic properties of the EuCl$_3$ monolayer, which constituted the primary objective of our research. To accomplish this, we employed an atomistic spin Hamiltonian based on a classical Heisenberg model, represented by Equation \ref{equ2}. By utilizing this model, we calculated the magnetic exchange coupling parameters, denoted as $J_{1}$, $J_{2}$, and $J_{3}$, which are essential components of the aforementioned equation. Our calculations yielded the following values for these parameters: $J_{1}$=\SI{-0.307}{\milli\eV}, $J_{2}$=\SI{-0.099}{\milli\eV}, and $J_{3}$=\SI{-0.405}{\milli\eV}. These values represent the strengths of the magnetic interactions within the EuCl$_3$ monolayer. Additionally, we determined the magnetic anisotropy energies (MAEs) of the monolayer, both in-plane $(E[100]-E[010])$ and out-of-plane $(E[100]-E[001])$. The calculated values for the in-plane and out-of-plane MAEs were found to be \SI{479}{\milli\eV} and \SI{573}{\milli\eV}, respectively. These values of magnetic anisotropy constants indicate that the major contribution to the total order parameter comes from the $z$-component of magnetic dipole moment.

\begin{figure}[h]
\centering
\includegraphics[scale=0.45]{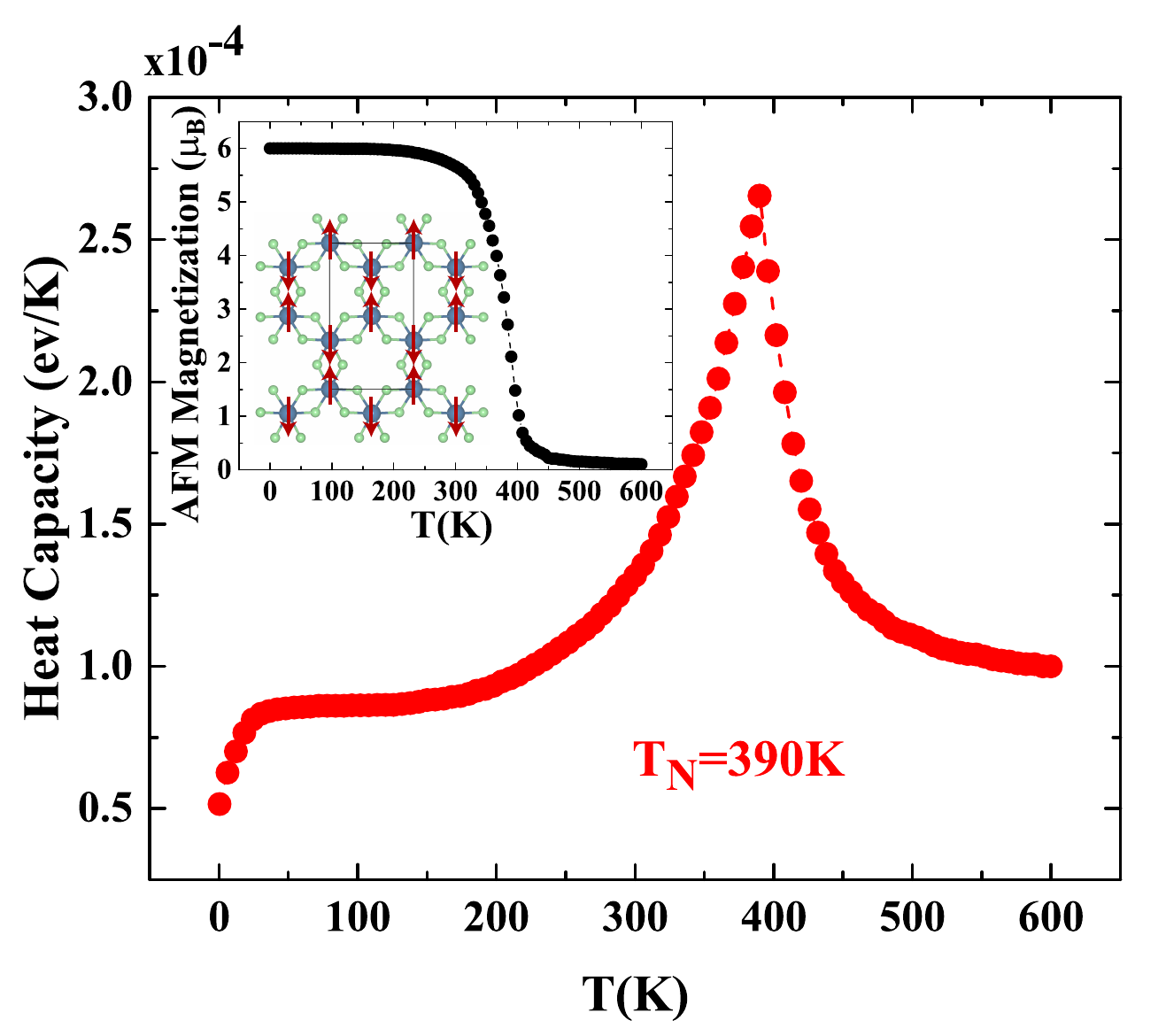} 
\caption{Variation of the heat capacity $\mathbf{C}$ as a function of temperature $T$. The inset shows the temperature dependence of AFM N\'{e}el magnetization and the schematic diagram depicts the ground state magnetic configuration of Eu atoms.}
\label{fig3}
\end{figure}

Fig. \ref{fig3} illustrates the temperature-dependent behavior of the order parameter $\mathbf{O}$ and heat capacity $\mathbf{C}$ of the EuCl$_3$ monolayer. The N\'{e}el order parameter $\mathbf{O}(T)$, representing the staggered magnetization, displays a distinct variation. The curve reveals that the ground state of the system is characterized by an antiferromagnetic (AFM) arrangement. As the temperature $T$ increases toward the critical region, a magnetic order emerges, and the order parameter reaches saturation at a value of $\mu\approx6.0$ within a sufficiently low-temperature regime. To identify the critical temperature of the system, we examine the variation of the heat capacity, represented by the curve $\mathbf{C}$ vs. $T$. The temperature corresponding to the peak position in this curve is recognized as the critical temperature. Our analysis indicates that the critical temperature, denoted as $T_{N}$, is determined to be \SI{390}{\kelvin}. This finding demonstrates that the magnetic order in the EuCl$_3$ monolayer can be sustained above room temperature.

\section{Conclusion}

In this study, we have thoroughly investigated the structural, electronic, and magnetic properties of the EuCl$_3$ monolayer, focusing on its potential applications in spintronics and magnetic devices. We found that the monolayer adopts a trigonal symmetry within the $P\bar{3}1m$ space group, with europium (Eu) atoms forming an octahedral coordination geometry coordinated by six chlorine (Cl) atoms. The ground state of the EuCl$_3$ monolayer exhibits a delicate balance between competing magnetic interactions, transitioning from a ferromagnetic to an antiferromagnetic N\'{e}el state when incorporating spin-orbit coupling (SOC) and electron-electron interactions through the PBE+SOC+U calculation. Regarding the electronic properties, our investigations showed that the EuCl$_3$ monolayer exhibited a semiconductor behavior with a band gap of \SI{1.75}{\eV} under the PBE functional and a significantly wider band gap of \SI{4.99}{\eV} when calculated with the HSE hybrid functional. The conducting behavior was found to be influenced by the $4f$ electronic states of the Eu atoms. Finally, through the temperature-dependent behavior of the order parameter and heat capacity, we found that the EuCl$_3$ monolayer exhibited an antiferromagnetic ground state with a critical temperature of \SI{390}{\kelvin}, indicating the potential for maintaining magnetic order above room temperature.

\begin{acknowledgments}
The computational resources are provided by T\"UBITAK ULAKBIM, High Performance and Grid Computing Center (TR-Grid e-Infrastructure), and the Leibniz Supercomputing Centre.  E. A.  acknowledges the Alexander von Humboldt Foundation for a Research Fellowship for Experienced Researchers. IO thanks to the Council of Higher Education (the CoHE) 100/2000 Program and to the Scientific and Technological Research Council of Turkey (TUBITAK) BIDEB-2214/A Program for providing doctoral scholarships. This research was supported by Aydin Adnan Menderes University Research Fund (Project Number: FEF-21032).
\end{acknowledgments}

\bibliography{bibliography}

\end{document}


\title{\textbf{Supporting Information: Emergent Antiferromagnetic Behavior in EuCl$_3$ Monolayer: A Comprehensive First-Principles Study Including Hubbard-SOC Interplay}}

\author{Nurcan Kutluk Kalkan}
\affiliation{Physics Department, Adnan Menderes University, 09100 Aydin,
Turkey}

\author{Ilkay Ozdemir}
\affiliation{Physics Department, Adnan Menderes University, 09100 Aydin,
Turkey}
\affiliation{Department of Physics, University of Antwerp, Groenenborgerlaan 171, B-2020 Antwerp, Belgium}

\author{Yusuf Y\"{u}ksel}
\affiliation{Physics Department, Faculty of Science, Tinaztepe Campus, Dokuz Eylul University, 35390 Izmir, Turkey.}

\author{\"{U}mit Ak{\i}nc{\i}}
\affiliation{Physics Department, Faculty of Science, Tinaztepe Campus, Dokuz Eylul University, 35390 Izmir, Turkey.}

\author{Johannes V. Barth}
\affiliation{Physics Department E20, Technical University of Munich (TUM), James Franck Strasse1, 85748, Garching (Germany)}

\author{Ethem Akt\"urk}
\email{ethem.akturk@adu.edu.tr}
\affiliation{Physics Department, Adnan Menderes University, 09100 Aydin,
Turkey}
\affiliation{Physics Department E20, Technical University of Munich (TUM), James Franck Strasse1, 85748, Garching (Germany)}

\date{\today}
\keywords{Hubbard-SOC interplay, Monte Carlo simulations, Magnetic Ground State, EuCl$_3$ monolayer}

\begin{abstract}

We present a comprehensive study on the electronic and magnetic properties of the EuCl$_3$ monolayer using first-principles calculations.  By taking into account the spin-orbit coupling (SOC) and the Hubbard effects,  we elucidate the influence of these interactions on the structural properties of EuCl$_3$ monolayer.  Comparing the lattice parameters obtained with the PBE+SOC and PBE+SOC+Hubbard effects reveals an increase in the lattice parameters when including the Hubbard effect.  In the absence of the Hubbard interaction, the magnetic ground state of the EuCl$_3$ monolayer exhibits a preference for a ferromagnetic (FM) configuration as determined by the PBE+SOC calculations.  However, the introduction of the Hubbard parameter leads to a shift in the magnetic ground state preference towards an antiferromagnetic (AFM) N\'{e}el state.  Based on the calculated energy values, Monte Carlo simulations are carried out to determine the N\'{e}el temperature $(T_N)$.  Our simulations yield a N\'{e}el temperature of \SI{390}{\kelvin} for the EuCl$_3$ monolayer, indicating the temperature at which the transition from a paramagnetic phase to an antiferromagnetic phase occurs.  These results highlight the importance of incorporating the SOC and the Hubbard effect in accurately describing the electronic and magnetic properties of the EuCl$_3$ monolayer. Our results contribute to a deeper understanding of the fundamental physics underlying the behavior of this intriguing two-dimensional material and provide insights into its potential applications in spintronics and magnetic devices.

\end{abstract}

\maketitle


\subsection{Calculation of the Effective Hubbard Parameter}
In this section, we present the details of the calculations conducted to determine the effective Hubbard-U parameter for the $f$-orbitals of Eu atoms. The DFT+U is a method that was proposed to improve the description of systems with strongly correlated $d$ and $f$ electrons, which are often inaccurately described by standard LDA and GGA functionals \cite{anisimov1991band}. Among the various variants of the DFT+U method available in VASP \cite{vasp1,vasp2}, we employed the simplified (rotationally invariant) approach as introduced by Dudarev \textit{et al.} \cite{dudarev1998electron} In this approach, an effective Hubbard parameter, denoted as $U_{eff}$, is utilized, where $U_{eff}$ is calculated as the difference between the Hubbard parameter $(U)$ and the Hund's exchange parameter $(J)$. The Hubbard parameter, $U$, represents the on-site Coulomb repulsion between electrons on the same atomic orbital, while the Hund's exchange parameter, $J$, accounts for the coupling of electrons with parallel spins within the same atomic shell. For our calculations, we set $J$ equal to 0, leading to $U_{eff}$ being equivalent to $U$ itself. To determine the effective Hubbard U term entering the expression of the functional, we employed a linear response approach, as introduced by Cococcioni and de Gironcoli \cite{cococcioni2005linear}. This method involves calculating both non-self-consistent (NSCF) $\chi_{IJ}^{0}=\frac{\partial N_{I}^{NSCF}}{\partial V_J}$ and self-consistent (SCF) $\chi_{IJ}=\frac{\partial N_{I}^{SCF}}{\partial V_J}$ response functions. These functions represent the system's response to localized perturbations in terms of non-interacting and interaction density responses. Using this response-function language, the effective interaction parameter $U$ associated to site $I$ can be found from: \cite{cococcioni2005linear}
\begin{equation}
U=\chi^{-1}-\chi_{0}^{-1} \approx \left( \frac{\partial N_{I}^{SCF}}{\partial V_I} \right)^{-1} - \left( \frac{\partial N_{I}^{NSCF}}{\partial V_I} \right)^{-1}
\end{equation}

\begin{figure}[h]
\centering
\includegraphics[scale=1.0]{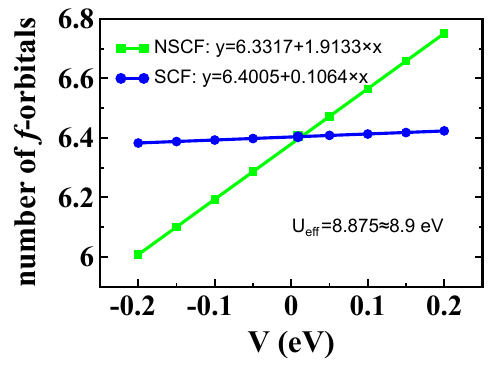} 
\caption{Linear fit of the number of $f$-orbital occupations on atomic site $1$ as a function of the additional potential $V$. The lines plotted with square and circle shapes correspond to the non-interacting (NSCF) and interacting (SCF) response functions, respectively.}
\label{figS1}
\end{figure} 

From Fig. \ref{figS1}, we then have:
\begin{equation}
U =\chi^{-1}-\chi_{0}^{-1} \approx \left( \frac{\partial N_{1}^{SCF}}{\partial V_1} \right)^{-1} - \left( \frac{\partial N_{1}^{NSCF}}{\partial V_1} \right)^{-1}
 = \frac{1}{0.1064} - \frac{1}{1.9133} = 8.875 \approx 8.9~ eV
\end{equation}

\subsection{Determining the Magnetic Ground State}
\begin{figure}[h]
\centering
\includegraphics[scale=0.5]{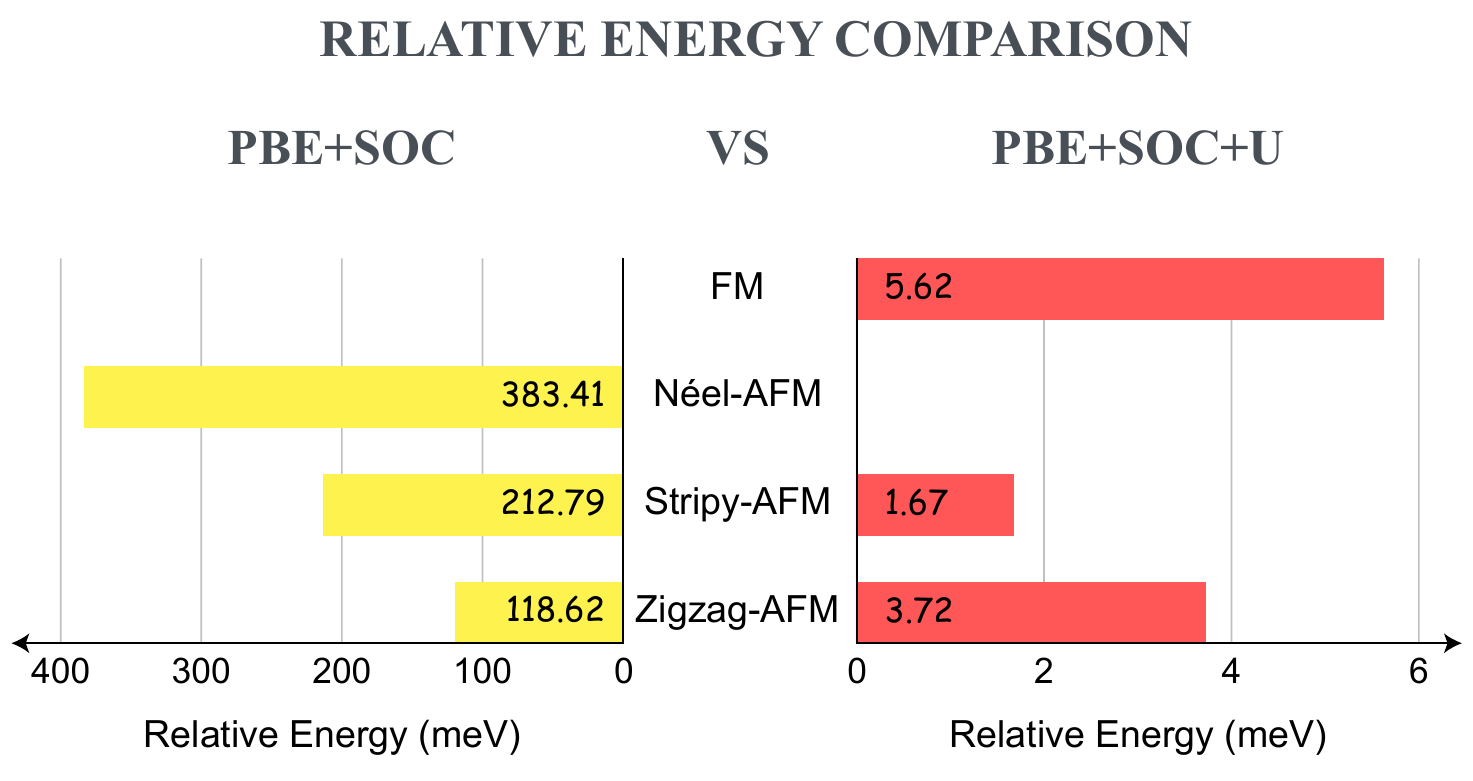} 
\caption{Comparison of relative energies between magnetic configurations calculated using the PBE+SOC and PBE+SOC+U methods.}
\label{figS2}
\end{figure}

In this section, we present a comparison of the calculated total energy values for various magnetic configurations, both without and with including the Hubbard U parameter. As shown in Fig. \ref{figS2}, it is evident that the EuCl$_3$ monolayer exhibits a preference for the ferromagnetic configuration as its magnetic ground state in PBE+SOC calculations. However, this preference shifts to the antiferromagnetic N\'{e}el state by introducing the effective Hubbard parameter.

\subsection{Structural Stability}
\begin{figure}[h]
\centering
\includegraphics[scale=1.0]{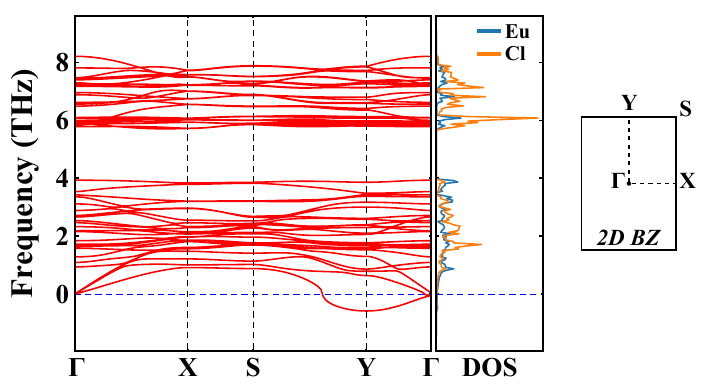} 
\caption{Phonon dispersion spectra and corresponding phonon projected density of states (PDOS) for the antiferromagnetic (AFM)-N\'{e}el EuCl$_3$ monolayer, calculated along the main symmetry directions of the two-dimensional (2D) Brillouin zone (BZ). The 2D BZ diagram is also provided.}
\label{figS3}
\end{figure}

In this section, we present our findings regarding the phonon frequencies throughout the Brillouin zone, as well as the phonon projected density of states of the AFM-N\'{e}el EuCl$_3$ monolayer. (See Fig. \ref{figS3}) Obviously, we observed predominantly positive phonon frequencies throughout the entire Brillouin zone, with the exception of the region around the $Y$-point. While this negative region might raise concerns about instability, we have carefully considered possible factors contributing to this behavior. One plausible explanation for the negativity might be the presence of inherent interactions, such as electron-phonon coupling or the effect of electron correlations localized in the $d$-orbitals of the Eu atoms. Our calculations does not include the Hubbard term for the $d$-orbitals of the Eu atoms since the VASP tool allows for the inclusion of Hubbard interaction only for one type of orbital for each atomic species. In the context of phonon calculations, the dynamical matrix is constructed using forces on atoms in the crystal structure to determine the phonon frequencies. The forces are sensitive to the electronic structure and bonding around each atom. While the Hubbard U term improves the description of the $f$-orbital electrons and their interactions, it may not be sufficient to fully capture the electronic effects on the neighboring $d$-orbitals. In some cases, the $d$-orbitals might also experience electron correlations or interactions with the $f$-orbitals. In short, if the $d$-electron correlations are not accurately accounted for in the DFT calculation (e.g., due to the lack of a Hubbard U term for the $d$-orbitals), this can lead to small negative frequencies in the phonon spectra. Despite this negativity, we do not believe that the system is inherently unstable; rather, it suggests the involvement of intricate physical interactions at play. It is worth noting that significant electron-phonon coupling could lead to phonon softening \cite{wallace1972thermodynamics}. Such a phonon softening phenomenon in the transverse acoustic phonon mode at low temperatures has been reported in the case of EuO, where it is attributed to potential spin-phonon coupling \cite{pradip2016lattice}.

\subsection{Electronic Properties of EuCl$_3$ Monolayer}
\begin{figure}[h]
\centering
\includegraphics[scale=0.8]{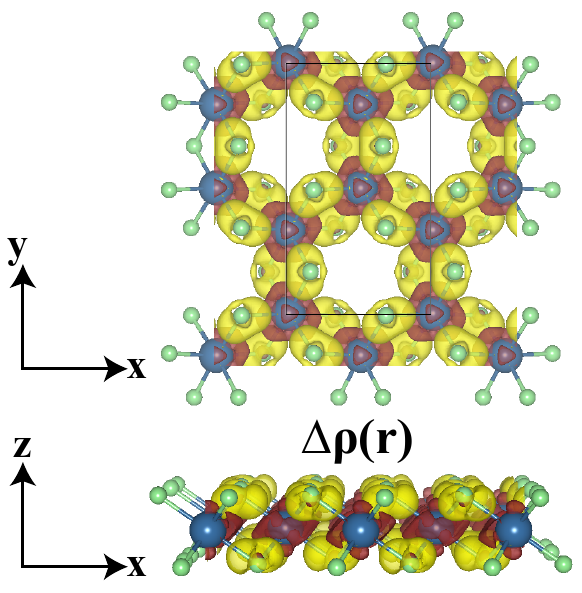} 
\caption{Isosurfaces of electronic charge density difference $\Delta\rho(r)$. Yellow and red isosurfaces indicate the positive and negative charge regions, respectively. Here, the isosurface value is set to \SI{0.005}{e \per\angstrom^3}.}
\label{figS4}
\end{figure}

In this section, we present the electronic properties of the EuCl$_3$ monolayer in its AFM-N\'{e}el ground state. The investigation begins with an analysis of the electronic charge density difference, denoted as $\Delta\rho(r)$. This quantity is obtained by subtracting the charge density contributed by the free Eu and Cl atoms at their respective atomic sites from the overall charge density of the entire system. Fig. \ref{figS4} visually presents the calculated $\Delta\rho(r)$ through isosurfaces, where the regions of positive and negative charge are represented in yellow and red, respectively. It is essential to clarify that the negative charge regions indicate a reduction in electronic charge density, signifying the direction of charge transfer. Conversely, the yellow isosurfaces correspond to regions of increased electronic charge density. This finding is consistent with the calculated Bader charges, which provide a quantitative measure of the electron distribution around each atomic nucleus within the EuCl$_3$ monolayer. Additionally, the agreement with their Pauli electronegativity values further supports the validity of our results.

\begin{figure}[h]
\centering
\includegraphics[scale=0.8]{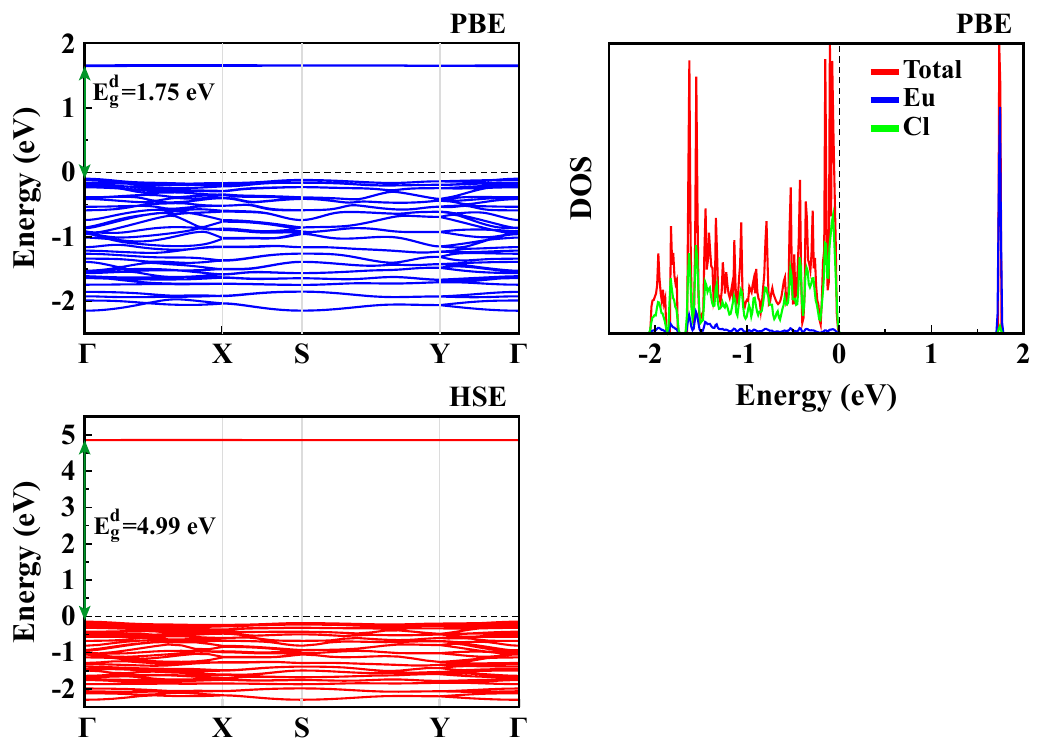} 
\caption{Electronic energy band structure of the EuCl$_3$ monolayer calculated within standard PBE fucntional and HSE hybrid functional along major symmetry directions of the two-dimensional Brillouin zone. The corresponding electronic density of states (DOS) obtained through PBE functional. The Fermi level is indicated by the black dashed line, serving as the reference for the energy scale.}
\label{figS5}
\end{figure}

Following the charge density difference analysis, we proceed to calculate the electronic energy band structure using both the standard PBE functional and HSE hybrid functional. Additionally, the electronic density of states (DOS) plot is obtained using the PBE functional. These plots are presented in Fig. \ref{figS5}, providing essential insights into the distribution of electronic states and the bandgap characteristics of the EuCl$_3$ monolayer.

\begin{figure}[h]
\centering
\includegraphics[scale=0.8]{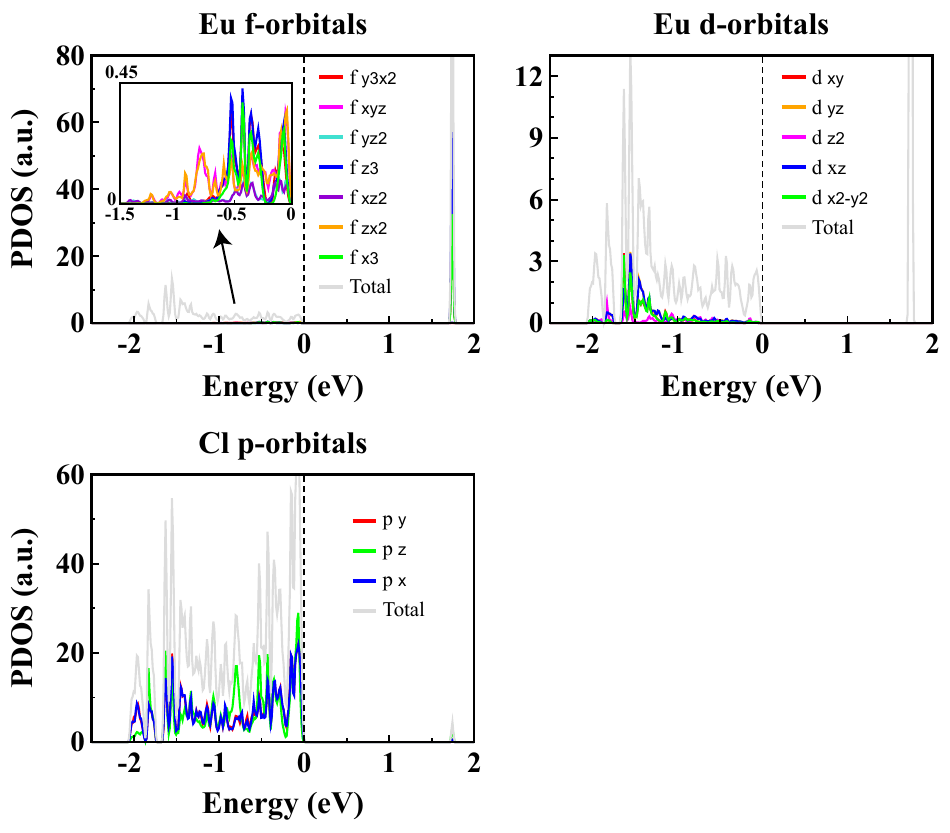} 
\caption{Partial density of states (PDOS) depicting contributions from $f$ and $d$ orbitals of Eu atoms and $p$ orbitals of Cl atoms in the EuCl$_3$ monolayer. The Fermi level is indicated by the black dashed line, serving as the reference for the energy scale.}
\label{figS6}
\end{figure}

Finally, to delve deeper into the nature of electronic states, we present the partial electronic density of states (PDOS) plot for the EuCl$_3$ monolayer in Fig. \ref{figS6}. The PDOS analysis provides a more localized perspective, attributing specific electronic states to individual atoms and orbitals.

\subsection{Magnetic Exchange Coupling Parameters}
In this section, we elucidate the process of deriving a set of equations that relate the energy of different magnetic configurations (see Fig. 2 in the main text), including ferromagnetic (FM), antiferromagnetic N\'{e}el (AFM-N\'{e}el), antiferromagnetic stripy (AFM-Stripy), and antiferromagnetic zigzag (AFM-Zigzag), through their mapping to the Hamiltonian presented in Equation 2 of the main text. This mapping results in the following system of equations:

\begin{equation}
 \begin{array}{lcl}
     E_{FM} &=& E_0 - (3J_1 + 6J_2 + 3J_1)\left|\vec{\mu}\right|^2  \\
   E_{Neel} &=& E_0 - (-3J_1 + 6J_2 - 3J_3)\left|\vec{\mu}\right|^2 \\
 E_{Zigzag} &=& E_0 - (J_1 - 2J_2 - 3J_3)\left|\vec{\mu}\right|^2   \\
 E_{Stripy} &=& E_0 - (-J_1 - 2J_2 + 3J_3)\left|\vec{\mu}\right|^2
 \end{array}
\end{equation}

Here, $\left|\vec{\mu}\right|$ represents the magnitude of the magnetic moment, equal to \SI{6.00}{\micro_B} (Bohr magnetons), and $E_0$ signifies the energy contribution originating from sources other than magnetic interactions. Solving this system of equations yields values for the four unknown parameters ($E_0, J_1, J_2, J_3$), which are presented in the table below.

\begin{table}[h]
\centering
\caption{Calculated exchange interaction values (meV)}
\begin{tabular}{cccc}
\hline \hline
 & $J_1$    & $J_2$    & $J_3$     \tabularnewline \hline
 &$-0.307$  &$-0.099$  &$-0.405$   \tabularnewline
\hline \hline
\label{tableS1}
\end{tabular}
\end{table}

The magnetic anisotropy constants, denoted as $k_y$ and $k_z$, are nothing but the energy differences $E[100]-E[010]$ and $E[100]-E[001]$, respectively, which are given in the main text.

\begin{acknowledgments}
The computational resources are provided by T\"UBITAK ULAKBIM, High Performance and Grid Computing Center (TR-Grid e-Infrastructure), and the Leibniz Supercomputing Centre.  E. A.  acknowledges the Alexander von Humboldt Foundation for a Research Fellowship for Experienced Researchers. IO thanks to the Council of Higher Education (the CoHE) 100/2000 Program and to the Scientific and Technological Research Council of Turkey (TUBITAK) BIDEB-2214/A Program for providing doctoral scholarships. This research was supported by Aydin Adnan Menderes University Research Fund (Project Number: FEF-21032).
\end{acknowledgments}

\bibliography{bibliography}